%% file: 00_elasticapp.tex
\documentclass[twocolumn]{wlscirep}
\usepackage[utf8]{inputenc}
\usepackage[T1]{fontenc}
\usepackage{indentfirst}
\usepackage[font=small,labelfont=bf, justification=justified, format=plain]{caption}
\usepackage{times}
\usepackage{graphicx}

\usepackage[switch]{lineno}

\usepackage[capitalise]{cleveref}   
\crefname{algocf}{Alg.}{Algs.}
\Crefname{algocf}{Algorithm}{Algorithms}

\usepackage[linesnumbered,ruled,vlined]{algorithm2e}

\usepackage{xcolor}
\usepackage{hyperref}
\hypersetup{
    colorlinks,
    linkcolor={red!100!black},
    citecolor={blue!100!black},
    urlcolor={blue!80!black}
}

\usepackage{siunitx}
\usepackage{amsmath}
\usepackage{bm} 
\usepackage[font={footnotesize}]{caption}

\input{_definitions.tex}

\title{Elastica++: A high-performance, multiphysics framework for large interacting assemblies of Cosserat rods}

\author[*,1,4]{Tejaswin Parthasarathy}
\author[*,1]{Seung Hyun Kim}
\author[*,1]{Songyuan Cui}
\author[1,2,3,$\dagger$]{Mattia Gazzola}

\makeatletter
\renewcommand\AB@affilsepx{, \protect\Affilfont}
\makeatother

\affil[1]{Department of Mechanical Science and Engineering, University of Illinois Urbana-Champaign, Urbana, IL, USA} 
\affil[2]{National Center for Supercomputing Applications, University of Illinois Urbana-Champaign, Urbana, IL, USA}
\affil[3]{Carl R. Woese Institute for Genomic Biology, University of Illinois Urbana-Champaign, Urbana, IL, USA}
\affil[4]{Nvidia}
\affil[$\dagger$]{Corresponding author: mgazzola@illinois.edu}
\affil[*]{These authors contributed equally to this work.}

\date{}

\begin{abstract}
Soft, slender structures are ubiquitous in natural and engineered systems, with broad application potential from biomimetic materials to soft robotics. However, there is a notable lack of computational tools that simultaneously preserve high-fidelity continuum rod mechanics, scale to large interacting ensembles, and remain flexible across diverse biophysical settings. Here we introduce Elastica++, an open-source, high-performance implementation of the Cosserat-rod model for large-scale simulations of slender-body dynamics. Elastica++ combines performance-oriented kernels with shared-memory parallelism to sustain teraflop-scale throughput despite complex discretization domains and physical interactions. The framework further interoperates with external numerical solvers, supporting efficient multiphysics workflows. We demonstrate robustness and breadth through case studies spanning passive nest-like metamaterials, collective active-matter dynamics, cilia carpets, soft magnetic microrobots, and schooling swimmers. Elastica++ thus provides a missing foundation for high-throughput studies of emergent behavior in interacting assemblies of elastic slender structures.
\end{abstract}

\begin{document}
\maketitle
\input{01_main}         
\input{02_results}      
\input{03_discussion}   
\input{04_methods}      

{\footnotesize \bibliography{elasticapp}}

\input{05_acknowledgements}   

\end{document}

%% file: _definitions.tex
\newcommand{\bv}[1]{\ensuremath{\bm{#1}}} 
\newcommand{\bt}[1]{\ensuremath{\mathbf{#1}}} 
\newcommand{\order}[1]{\ensuremath{\mathcal{O}(#1)}}

\newcommand{\grot}{\ensuremath{\textup{SO}(3)}}

\newcommand{\elastica}{\textit{Elastica++}}
\newcommand{\etal}{\textit{et al.}}

\renewcommand{\emph}[1]{\textit{#1}}

%% file: 01_main.tex
Soft slender structures are ubiquitous in nature, fulfilling diverse functions ranging from passive structural support~\cite{hansell2000bird,weiner2020mechanics} to active manipulation~\cite{aydin2019neuromuscular,zhang2019modeling,marvi2014sidewinding}, as illustrated by microcilia that drive fluid motion~\cite{dong2020bioinspired,gu2020magnetic}, the densely-packed fiber architecture of muscular hydrostats~\cite{tekinalp2024topology}, and aerial, web-like assemblies of slender threads~\cite{bhosale2022micromechanical}.
These natural systems have inspired interdisciplinary efforts to translate biological organizational principles into engineered platforms, with applications in soft robotics, biomimetic materials, metamaterials, flexible electronics, and biomedical devices~\cite{rus2015design,wegst2015bioinspired,bertoldi2017flexible,rogers2010materials,cianchetti2018biomedical,li2019origami,li2015fluidic,tao2023engineering}.
To capture the large deformations and strong nonlinearities of soft slender bodies, the Cosserat rod theory has been widely explored~\cite{CosseratThorieDC,ericksen1957exact,antman1974kirchhoff,simo1986three,antman2005nonlinear,altenbach2013cosserat}, providing a geometrically exact one-dimensional continuum framework that accommodates all practically relevant deformation modes, including bending, twist, shear, and extension.
More recently, broader applications of geometrically exact rod models~\cite{lazarus2013continuation,zhang2019modeling} have bridged the interpretation of natural phenomena and the design of engineered systems~\cite{till2019real}.

Yet, many of the most compelling natural behaviors, including collective motion~\cite{bar2020self,doostmohammadi2018active}, environmental reconfiguration~\cite{gu2020magnetic,weiner2020mechanics,chen2018envshell}, and versatile functionality~\cite{lu2018bioinspired,shih2023hierarchical}, cannot be explained by deformation alone.
They arise from interaction-dominated physics, where multiple slender bodies couple with one another and with surrounding fluids, surfaces, contacts, and external fields~\cite{jawed2014coiling}.
In many settings, resolving these multi-rod interactions and environmental couplings poses a greater bottleneck than describing the kinematics of an individual rod~\cite{harmon2009asynchronous}.
A unified, consistent computational framework that can simulate large ensembles of interacting rods under heterogeneous physics, while remaining tractable for rapid prototyping, design iteration, parameter exploration, and learning-oriented workflows, remains elusive.
Existing high-fidelity approaches, such as full three-dimensional finite-element methods for continuum elasticity~\cite{bathe2006finite,bucalem1995locking,felippa2005unified}, can provide accurate simulations in complex settings, but become prohibitively expensive for dense fibrous assemblies.
Conversely, highly lumped models, while common in robotics and computationally efficient~\cite{webster2010design,renda2018discrete}, may miss key attributes of distributed elastic, geometric, and contact mechanisms~\cite{wang2024sensing,marvi2012friction}.
This gap points to the need for a modeling framework that preserves the essential mechanics of slender-body systems while supporting efficient computation, flexible problem configuration, and extensible coupling to diverse physical interactions.

Responding to these challenges, we present \elastica{}, an open-source, performance-oriented C++ framework implementing a discrete Cosserat rod model~\cite{gazzola2018forward,pyelastica} at scale.
Redesigned for shared-memory parallelism and many-rod simulations with heterogeneous physics, \elastica{} exposes a composable set of primitives for contact, friction, internal actuation, external loads, and optional coupling to external numerical tools such as fluid solvers~\cite{sopht}.
The implementation pairs template-driven customization of geometry, constitutive laws, and interaction modules with aggressive compute optimizations---cache-friendly memory layouts, vectorization, and task-based threading~\cite{pheatt2008intel}---yielding significant speed-ups and enabling far larger filament counts than previous frameworks~\cite{gazzola2018forward}.
We demonstrate these capabilities through a diverse suite of scenarios including fibrous granular packings, dense active filament suspensions, magnetically driven cilia arrays, and fluid-structure schooling, establishing the framework's robustness and scalability for emergent collective dynamics.

%% file: 02_results.tex
\section*{Results}

\begin{figure*}[ht!]
    \centering
    \includegraphics[width=\linewidth]{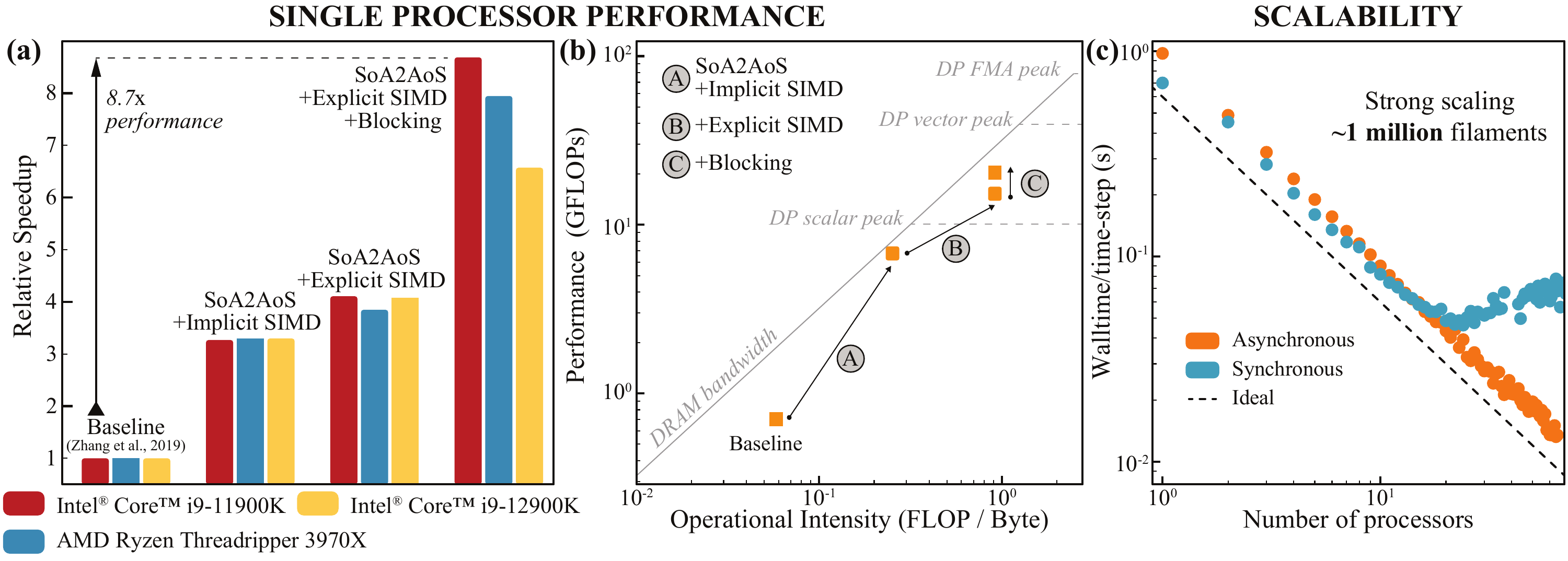}
    \caption{
    Performance improvements in \elastica{} via various HPC strategies. 
    \textbf{(a)} Ablation study on different processors: relative speed-ups on a single processor obtained cumulatively via (i) SoA2AoS data layout transformations, (ii) explicit (programmer-defined) SIMD, and (iii) data aggregation and blocking, compared against the unoptimized baseline. 
    \textbf{(b)} Roofline analysis of the in-place SO(3) rotation kernel with successive application of optimization methods. 
    All data are measured on an Intel Core i9--11900K CPU via Intel VTune Advisor. 
    The gray lines indicate the theoretical double-precision peak performance and memory bandwidth of the processor, accounting for effects of vectorization and fused-multiply-add (FMA) operations.
    \textbf{(c)} Strong scaling of a parallel \elastica~simulation with $1024^2$ filaments with and without thread synchronization after each symplectic integration stage.
    Experiments are performed on the Expanse supercomputing cluster~\cite{expansesdsc} with up to 64 AMD EPYC 7742 processors.
    }\label{fig:block_result}
\end{figure*}

\subsection*{Performance-oriented software design}
Naively, numerical algorithms for Cosserat rod dynamics (see Methods) involve $\order{n}$ complexity in both computation and memory, rendering the algorithm intrinsically memory bound which makes optimal hardware utilization a difficult task.
\elastica{} leverages parallel computing paradigms including data layout transformations (e.g., "struct of arrays" to "array of structs" / "SoA2AoS") and explicit Single Instruction Multiple Data (SIMD), as well as block memory aggregation where data structures across multiple rods are grouped to facilitate data locality, achieving significant single-core performance gain (\cref{fig:block_result}a).
Compared with a baseline implementation without these optimizations, \elastica{} achieves up to $8.7\times$ speed-up on a single processor across different CPU architectures.
Furthermore, we evaluate the machine utilization via a roofline analysis of the in-place \grot~rotation (\cref{fig:block_result}b), the most computationally intensive kernel in~\elastica{}.
We measure the performance (in GFLOPS) with respect to operational intensity (OI), defined as the number of effective floating-point operations per byte of off-chip memory loaded. 
The collection of optimizations imposed is shown to increase the OI to $\sim 1$ FLOP/Byte, yielding significant performance improvements from $\sim 0.7$ GFLOPS to a sustained $\sim 20$ GFLOPS.
While the kernel execution is still memory bound, the prescribed optimizations enable successful extraction of 95\% single-core peak performance on the tested processor (Intel Core i9--11900K), prompting further speed-ups via multi-core thread-level parallelism.
Indeed, we leverage Intel One-API Threading Building Block (TBB)~\cite{pheatt2008intel} to investigate the scalability of \elastica{} on multiple processors, via a dynamical simulation with $1024^2 = \num{1048576}$ filaments. 
We time-integrate the rod dynamics via a multi-stage symplectic (Verlet) algorithm, wherein each stage requires enforcing kinematic constraints. 
\elastica{} can be configured to apply thread synchronization after each stage (synchronous) or to defer synchronization to the end of the full time-step (asynchronous), the latter being more efficient but also restrictive in scenarios involving rod interactions.
\Cref{fig:block_result}c shows the strong scaling of both protocols, from which we conclude that the asynchronous protocol yields near-optimal scaling behavior, while the synchronous simulation deviates at high numbers of processors due to serial synchronization code ($\sim 6\%$ of the total time). 
For additional details on the performance optimizations and scaling, the reader is referred to supplementary materials, Section 3.

Building on this performance characterization, each subsections below present several physics-based numerical studies organized by application cases.  

\subsection*{Fibrous granular materials}
\input{case_studies/nest}

\subsection*{Active matter}
\input{case_studies/active_matter}

\subsection*{Magnetized filament assemblies}
\label{sec:magnetic-case}
\input{case_studies/magnetism}

\subsection*{Third-party software integration}
\input{case_studies/flow_coupling}

%% file: case_studies/nest.tex
\begin{figure*}[ht!]
    \centering
    \includegraphics[width=0.95\textwidth]{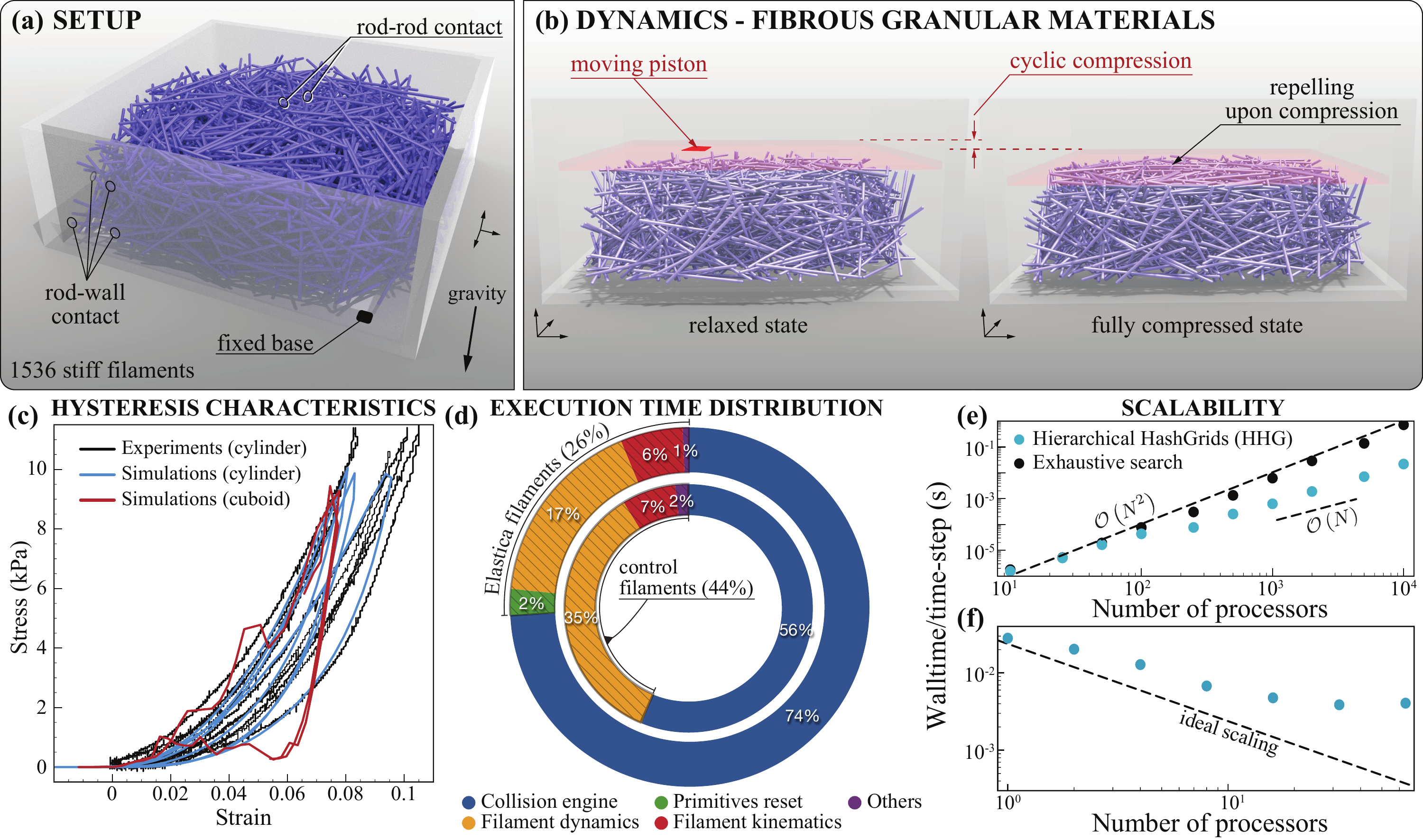}
    \caption{
    Fibrous granular materials. 
    \textbf{(a)} Physical setup of a random packing of $\num{1536}$ filaments (length $L$, aspect ratio $AR=31$) in a top-open cuboidal enclosure of size $3.34L$. 
    \textbf{(b)} A piston, modeled as a moving solid boundary, cyclically compresses the packing.
    \textbf{(c)} Bulk mechanical responses depicted by quasi-static stress-strain cycles for simulations and control experiments. 
    \textbf{(d)} Execution time distribution for single-processor \elastica~and control simulations. 
    \textbf{(e)} Time-step average broad-phase collision time via exhaustive search and hierarchical hash-grid (HHG), scaled by the number of filaments $N$. 
    \textbf{(f)} Shared memory strong scaling of the \elastica~simulation with up to 64 processors. 
    }
    \label{fig:nest}
\end{figure*}

We demonstrate {\elastica}'s quantitative capability for simulating disordered packing of unbounded, semi-flexible fibers, as seen in twig-based bird nests~\cite{hansell2000bird} and unwoven textiles~\cite{poquillon2005experimental}. 
Here, we simulate an expanded version of the experimental and numerical system in our previous work Bhosale \etal~\cite{bhosale2022micromechanical}, increasing the number of filaments from $460$ to $\num{1536}$ at fixed packing density, to capture the bending-driven nonlinear bulk behavior that arises from evolving micro-mechanical contacts.
The simulation entails filaments placed randomly inside the enclosure while a moving piston, modeled as a solid boundary, cyclically compresses the packing (\cref{fig:nest}b,c). 
We account for rod interactions via normal force and friction, with dynamical parameters and compression protocol kept consistent with the control study~\cite{bhosale2022micromechanical}.

We time-march the system in \elastica~and collect the quasi-static loading response alongside prior experiments and simulations (\Cref{fig:nest}c).
Discrepancies are due to wall-boundary configuration: compression drives filaments into cuboid corners; nevertheless, the up-scaled simulation retains key qualitative features, including distinct loading and unloading stress-strain branches that indicate \textit{hysteresis}.
As in prior work, this history dependence is attributed to non-viscous dissipation from inter-filament friction~\cite{bhosale2022micromechanical}. 

\Cref{fig:nest}d provides execution-time profiling of each kernel for both the present and control studies~\cite{bhosale2022micromechanical}.
In both cases, the collision engine (blue), which detects and resolves rod-rod and rod-boundary contact with friction, dominates runtime.
We note that both studies report a similar average coordination number during compression, $\langle z \rangle \approx 7$~\cite{bhosale2022micromechanical}, yet the previous implementation in Bhosale \etal~\cite{bhosale2022micromechanical} spent $44\%$ of its runtime solving the rod equations, compared with $26\%$ in {\elastica}. 
This reduction reflects significant speed-ups in {\elastica}'s core kernels and indicates collision resolution is now the primary bottleneck.
The main acceleration in the collision engine is twofold: (1) efficiently detecting and pruning via a hierarchical hash-grid (HHG) algorithm~\cite{eitz2007hierarchical} ($\order{N^2}\rightarrow\order{N}$), and (2) reducing non-coalesced sparse memory access caused by irregular contact locations (\cref{fig:nest}e).
Additional HHG implementation details are provided in supplementary materials Section 2.

Finally, \Cref{fig:nest}f shows strong scaling across multiple processors, restricted to synchronous execution because of collision resolution constraints. 
The observed behavior corresponds to an application with a $15\%$ serial portion, consistent with prior single-node multi-core studies of similar discrete element methods~\cite{berger2015hybrid,yan2019comparison}. 
This serialization is largely comprised of sequential steps of hash-grid implementation, where fine-grained data access limits parallelization. 
Given the high contact throughput implementation, GPU acceleration of the collision engine~\cite{fang2021chrono} is an immediate direction for future work.

%% file: case_studies/active_matter.tex
Next, we examine the inverse problem where active collectives interact within a passive environment, scaling to larger ensembles over longer time horizons.
This configuration reflects growing research of emergent behavior in \textit{self-propelled rods} and particles, from motile bacterial suspensions~\cite{wu2000particle} and actin cytoskeleton in biological cells~\cite{banerjee2020actin} to liquid crystals~\cite{bisoyi2021liquid} and synthetic Janus colloids~\cite{vinze2024self}.
Recently, significant inroads were made to understand the collective behaviors, such as clustering and synchrony, using over-damped Langevin simulations, where each filament is modeled as a chain of connected beads, with Brownian motion and stochastic interaction~\cite{isele2015self,anand2020conformation,he2025spontaneous}.
To extend for larger continuum description, the Cosserat rod model is an attractive alternative which tracks internal orientations through directors.
Here, we model a subclass of dry (absence of surrounding fluid) active-matter~\cite{bar2020self,doostmohammadi2018active}, where biologically accurate, short-ranged repulsive interactions govern emergent physics.

\begin{figure*}[ht!]
    \centering
    \includegraphics[width=\textwidth]{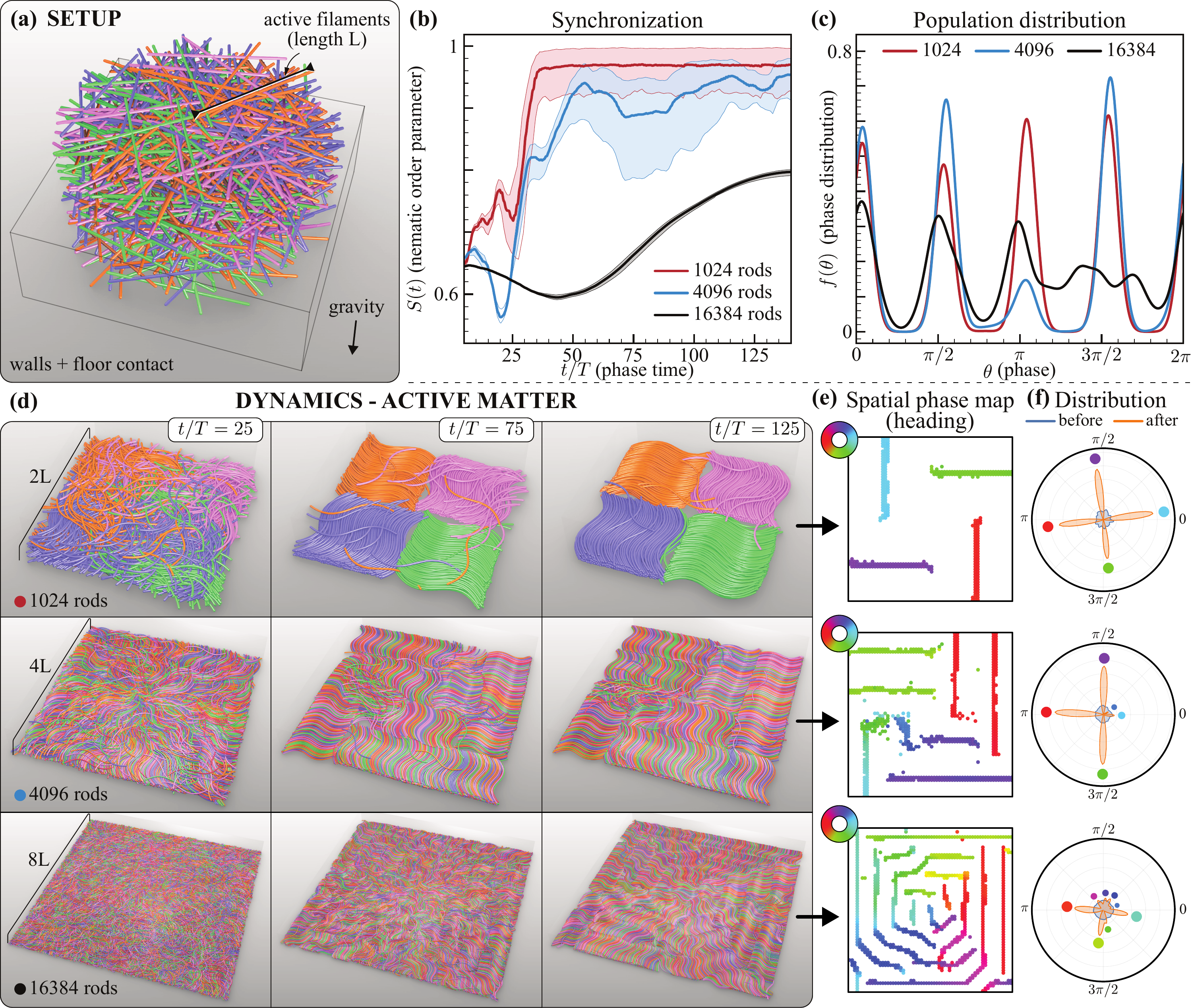}
    \caption{
        Active matter.
        \textbf{(a)} Physical setup illustrating initial random packing of sticks in a top-open cuboidal enclosure.
        \textbf{(b)} Evolution of scalar nematic order parameter $S(t), \bar{S}(t)$ for the system with 1024, 4096, 16384 filaments.
        \textbf{(c)} Polar probability distribution averaged over the 125th time period.
        \textbf{(d)} Simulation snapshots of rod clustering aggregation at different scales: 1024, 4096, and 16384 filaments, at $\tfrac{t}{T} = 25, 75, 125$ respectively.
        \textbf{(e)} Spatial map of filament centers, colored by polar angle of filament orientations. 
        \textbf{(f)} Distribution of orientations on the phase map, with indicator of an aggregated subpopulation.
    }
    \label{fig:active_matter}
\end{figure*}

We introduce a simulation with a random packing of $n_{\textrm{rods}} = 1024$ filaments (length $L$) in a cuboidal box (side-length $2L$), shown in \Cref{fig:active_matter}a.
The filaments settle under gravity, then self-activate following a bio-inspired snake-locomotion model by Zhang \etal~\cite{zhang2019modeling}, prescribed via internal (net-zero) muscular torques as a traveling wave with period $T$ and wavelength $L$.
In this study, we focus on the emergent long-term behavior of the collective system, influenced by a mixture of elasticity, self activation, frictional contacts, and boundary confinement.
We first show snapshots at three time instants in the columns of \Cref{fig:active_matter}d.
The filaments, initially random in both positions and orientations, are observed to \textit{nematically} aggregate towards the domain boundaries, similar to an investigation on colloidal rods~\cite{wensink2008aggregation}.
As time progresses, the boundary aggregations organize into four \textit{smectic} groups with synchronous motion and distinct orientations, each occupying a spatial quadrant.
We rationalize this behavior by noting that strong self-propulsion~\cite{bar2020self} lets filaments overcome friction and crawl over one another.
In such situations, the system effectively behaves as quasi-two-dimensional~\cite{ginelli2010large,bar2020self}, leading to the characteristic formation of dynamic nematic bands in which rods align parallel to their long axes.
Due to the high filament density and commensurate boundary dimensions, we further observe the four layered, synchronous substructures.

We quantify filament orientation by $\theta_{j} (t)$, $j = 1,\ldots,n_{\textrm{rods}}$, defined as the polar angle of the tail-to-head vector projected onto the frictional plane.
This further enables computation of the time-dependent, global, axis-aligned nematic order parameter $S(t) =  \langle |\real{\exp(i 2\theta_j)}| \rangle_{j}$ and its periodic moving average $\bar{S}(t) = \int_{t-T/2}^{t+T/2} S(\tau) d\tau$, where $\left\langle \cdot \right\rangle$ represents the ensemble expectation.
We show the time evolution of $\bar{S}(t)$ as a solid red curve in \Cref{fig:active_matter}b, surrounded by bands representing the extent of oscillations in $S(t)$ due to snake motion.
As the simulation progresses, $\bar{S}(t)$ approaches unity and stabilizes, indicating a fully synchronized, axis-aligned, nematically ordered system.
We then track subpopulations by reporting the polar orientation probability distribution $f(\theta)$, averaged over the 125th time-period, as a red curve in \Cref{fig:active_matter}c.
Four distinct peaks at $\theta = 0, 0.5\pi, \pi, 1.5\pi$ correspond to the four clusters in \Cref{fig:active_matter}d.

Lastly, we upscale the number of filaments while retaining filament density to test whether the collective behavior is preserved.
We demonstrate snapshots for $n_{\textrm{rods}} = 4096$ and for $n_{\textrm{rods}} = 16384$ (\cref{fig:active_matter}d), and plot the corresponding time evolution of synchrony $\bar{S}(t)$ in \Cref{fig:active_matter}b (blue and black, respectively).
The orientation probability distributions in \Cref{fig:active_matter}c again show subpopulation aggregation with a characteristic smectic pattern within each cluster.
Compared with the $n_{\textrm{rods}} = 1024$ case, scaled-up systems take significantly longer to reach and stabilize nematic order (\cref{fig:active_matter}b), and the spatial patterns become more complex (\cref{fig:active_matter}e,f), with layered smectic grain boundaries.
The supplementary video shows that ordering begins at contact boundaries and propagates inward, while regions enclosed by stabilized clusters require much longer to resolve competing orientations, suggesting slow relaxation and metastable states.
Overall, this case combines fine-grained biophysical modeling with flexibility and computational scale, offering a platform to test and refine theories of emergent micro- and macro-scopic organization in active matter.

%% file: case_studies/magnetism.tex
\begin{figure*}[ht!]
    \centering
    \includegraphics[width=\textwidth]{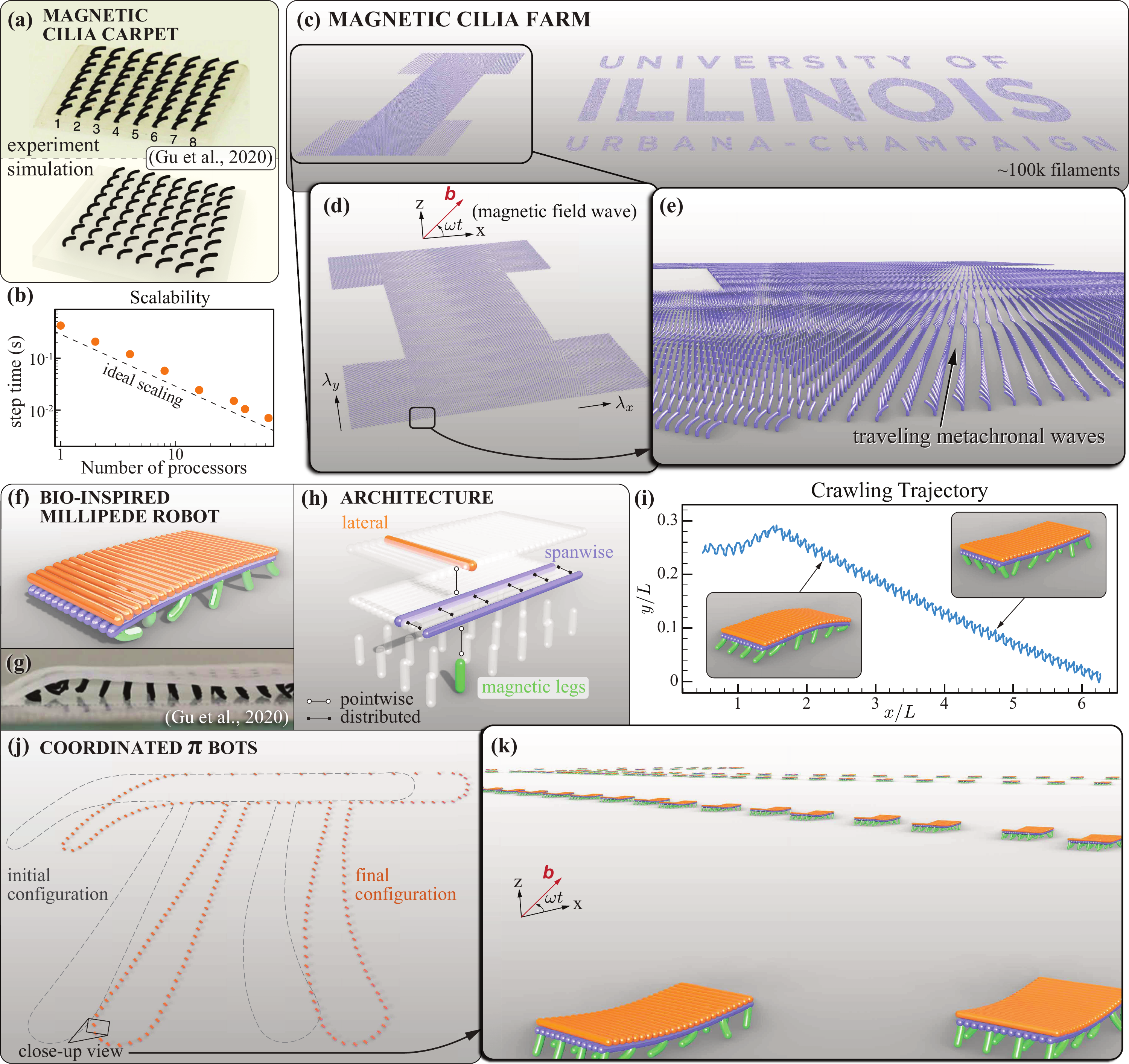}
    \caption{
        Magnetized filament assemblies.
        \textbf{(a)} Experiments~\cite{gu2020magnetic} and one-to-one simulations of an $8\times 8$ carpet of magnetized cilia.
        \textbf{(b)} Strong scaling of simulation step time with processor count.
        \textbf{(c, d)} Up-scaled carpet of $\sim \num{100000}$ cilia forming the University of Illinois Wordmark.
        \textbf{(e)} The metachronal wave pattern at scale.
        \textbf{(f, g)} Experimental~\cite{gu2020magnetic} and numerical realization of crawling millipede robots, driven magnetically via metachronal waves.
        \textbf{(h)} Layered architectural assembly of the \elastica~millipede using different types of connected filaments.
        \textbf{(i)} Trajectory of millipede motion on plane.
        \textbf{(j, k)} Coordinated arrangement of $224$ identical millipede bots in the shape of ``$\pi$'', crawling over 10 body lengths over $120$ magnetic cycles.
    }
    \label{fig:magnetism}
\end{figure*}

Next, we turn from internally actuated rods to magnetically driven filaments, motivated by programmable cilia carpets and coordination~\cite{gu2020magnetic,dong2020bioinspired,zhang2022metachronal}.
Each filament carries a prescribed body-frame magnetization that reacts to a time-varying external field via a magnetic torque (see Methods).
In \elastica{}, this forcing is added as a modular layer atop the elastic rod, frictional contact, and body-force components from the preceding case studies, keeping ablation and coupling analyses tractable.
We illustrate this through two applications: large-scale simulation of magnetic cilia carpet, and architected multi-component magnetic ``millipede'' soft robots (\cref{fig:magnetism}).

Inspired by cilia carpet in microfluidic applications\cite{gu2020magnetic,tekinalp2025self}, we simulate collections of ciliated filaments following beating motions under the oscillatory magnetic field.
Each filament cycles through stroke and recovery with spatially shifted timing, producing macroscopic \textit{antiplectic metachronal} waves (\cref{fig:magnetism}a).
We then scale the same construction to a ``cilia farm'' of $\num{111797}$ filaments (\cref{fig:magnetism}c-d), prescribing distinct wavelengths along $x$ and $y$ so the resulting wave is two-dimensional and travels diagonally across the domain (\cref{fig:magnetism}e).
Using asynchronous time integration, we achieve near-perfect strong scaling (\cref{fig:magnetism}b), demonstrating {\elastica}'s scalability for large, field-driven filament ensembles without distributed memory.

Turning to the swarm-microbot problem, we provide a computational realization of a magnetic ``millipede''~\cite{lu2018bioinspired,gu2020magnetic,venkiteswaran2020tandem,dong2020bioinspired}, experimentally studied but not previously instantiated \emph{in-silico} (\cref{fig:magnetism}f,g).
Three heterogeneous components comprise the bot (\cref{fig:magnetism}h): magnetized legs (green) and non-magnetized structural backbone filaments---spanwise (violet) and lateral (orange).
Components of different types are connected point-wise, while backbones of the same type are joined using lateral, distributed connections~\cite{tekinalp2024topology}.
Under a cyclic external field, the periodically varying magnetic torques on the legs produce metachronal waves, driving forward-crawling locomotion (\cref{fig:magnetism}i).
We then arrange $224$ such bots ($\num{13888}$ filaments total) in the shape of ``$\pi$'' (\cref{fig:magnetism}j,k); the swarm covers $> 10$ body lengths over $\sim 120$ magnetic cycles while preserving its formation, confirming good dynamical and numerical stability.
This opens opportunities for further study compliant designs and coordinated microbots actuated by internal and external mechanisms.

%% file: case_studies/flow_coupling.tex
\begin{figure*}[ht!]
    \centering
    \includegraphics[width=\textwidth]{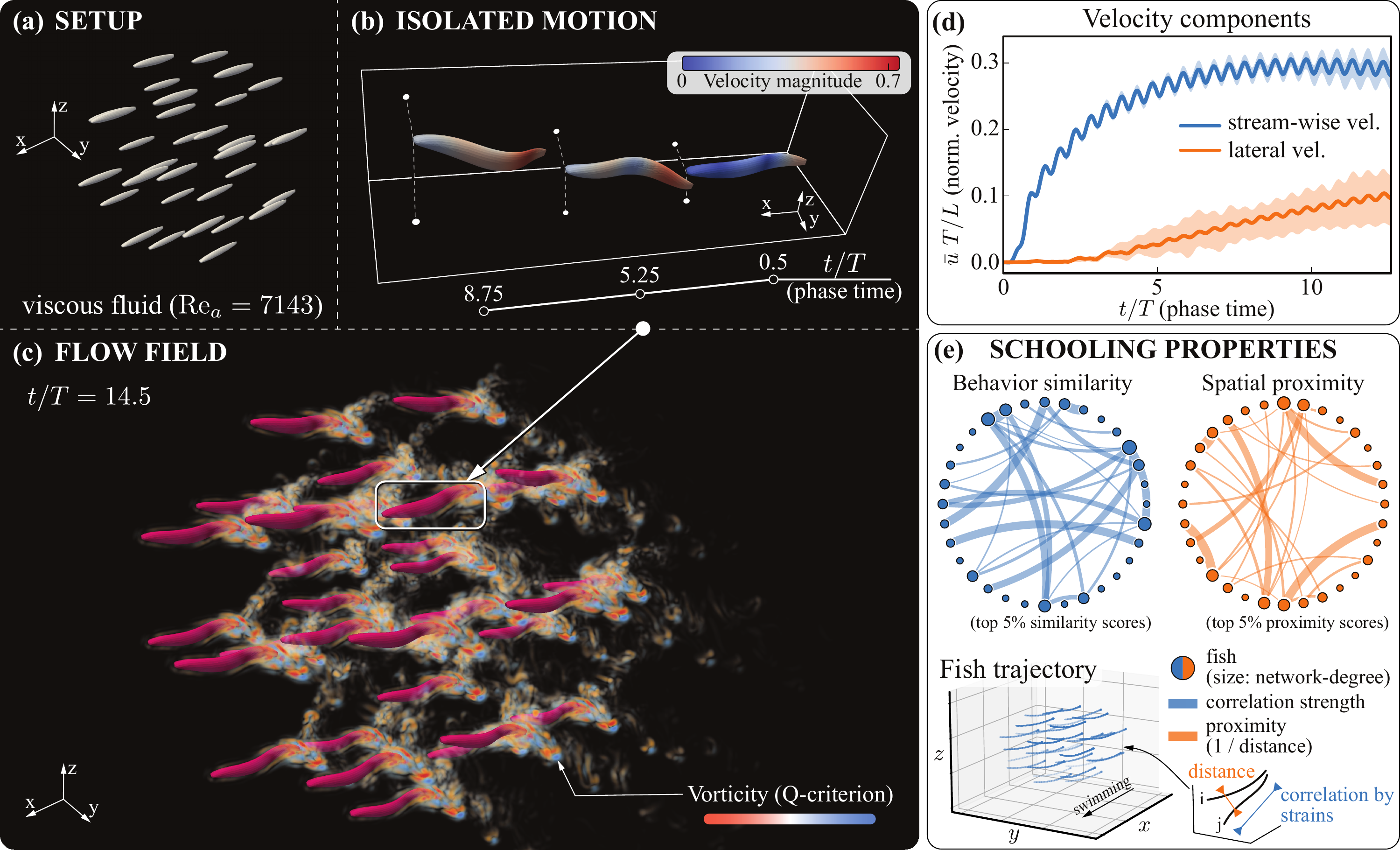}
    \caption{
        Schooling of anguilliform swimmers.
        \textbf{(a)} Simulation setup of $32$ anguilliform swimmers arranged in a uniformly-spaced unstructured lattice structure, immersed in a viscous, quiescent fluid ($\textrm{Re}_a := L^2 / (T\nu_f)=\num{7143}$, where $\nu_f$ is the fluid kinematic viscosity.).
        \textbf{(b)} The motion of a single swimmer at several time instants, colored by instantaneous velocity magnitudes.
        \textbf{(c)} Snapshot of the schooling swimmers and the flow field at $\tfrac{t}{T} = 14.5$, rendered using Q-criterion.
        \textbf{(d)} Time evolution of the streamwise (blue) and lateral (orange) velocity components, illustrated by both the average (solid lines) and ranges (shaded regions) across all swimmers.
        \textbf{(e)} Statistics of schooling properties based on pairwise analysis between swimming trajectories and deformation similarity.
        (blue) network represent the deformation similarity of the swimmers, measured by the correlation of bending fluctuations.
        (orange) network represent the spatial proximity, measured by the time-averaged center-of-mass separation.
        Bottom panel shows the center-of-mass trajectories for all swimmers over $14.5T$, with net motion in $x$ and drift in $z$.
    }
    \label{fig:fish_swimming}
\end{figure*}

To expand the scope of {\elastica} to more complex physical problems and broader application domains, its software architecture is built around modular design and flexible coupling protocols, enabling seamless integration with external software without compromising modeling fidelity and computational efficiency.
Here, we demonstrate this capability through a fluid-structure interaction (FSI) case study where we examine the swimming of a school of anguilliform swimmers (fishes), motivated by an aquatic propulsion in soft robotics and coordinated schools~\cite{carling1998self,bergmann2011modeling,weber2020optimal,bergou2010viscousflowrod}.

The setup closely follows previous work by Tekinalp \etal~\cite{tekinalp2025self} where an individual swimmer is modeled as a Cosserat rod (length $L$, elliptical cross-sections) in {\elastica}, and a third-party fluid solver~\cite{sopht} is used to incorporate the flow field. Swimmers implemented in {\elastica} interact with the fluid solver bi-directionally using the immersed boundary method.
While the soft-rod dynamics are handled in {\elastica}, the computationally-intensive numerical tasks for fluid-interaction, such as flow dynamics, immersed boundary interpolations, and boundary conditions (no-slip, no-penetration) are computed externally (supplementary materials, Section 4).
This enables a large-scale multi-physics problem with a school of $32$ fishes inside the flow domain, set to be $12L \times 8L \times 8L$ at a spatial resolution of $960 \times 640 \times 640$.
Fish-like swimming gaits are prescribed with traveling flexural wave (period $T$) activating internal torsional moments, while allowing compliance of the soft body to interact with the surrounding fluids.
For consistency with the work by Tekinalp \etal~\cite{tekinalp2025self}, we set $\textrm{Re}_a = \num{7143}$ and time march the simulation for $14.5T$, with the final snapshot of the swimmers and flow field shown in \Cref{fig:fish_swimming}c.
We note the emergence of coherent streets of vortices in the wake of each swimmer, visualized via the Q--criterion with the orange and blue regions representing vorticity and strain-dominated regions, respectively.

To analyze variability in fluid-driven deformation, we examine the swimming and schooling statistics and efficiency by tracking the center-of-mass position and velocity of each swimmer.
The velocity is decomposed into streamwise (forward, along the $x$-direction) and lateral ($yz$-plane) components.
We report the time evolution of both speeds, normalized by the characteristic velocity scale $L/T$, via their average (solid lines) and ranges (shaded regions) across all swimmers in \Cref{fig:fish_swimming}d.
We observe a consistent forward speed across all swimmers but significant variations in the lateral speed.
In particular, the lateral drift and its variability are attributed to a combination of hydrodynamic (in-)stability, boundary effects, and the interaction between swimmers by the vortical structures.
On the other hand, the normalized forward speed, while not exhibiting such significant variations, stabilizes around $0.3$, slightly lower than previously reported values for a single swimmer in Tekinalp \etal~\cite{tekinalp2025self}.
The discrepancy would be reduced by increasing the fluid grid resolution, but the steep increase in computational cost required by the third-party solver places this beyond the scope of the present work.


Finally, \Cref{fig:fish_swimming}e summarizes deformation similarity and the influence of fluidic effects within the schooling structure, which are not captured by bulk swimming speeds alone.
All fishes follow similar deformation patterns under identical actuation, but the vortical wake behind each fish impact trailing swimmers, leading to second-order variabilities.
To assess such variations in swimming cycles relative to spatial proximity, we performed two pairwise analyses: the correlation of bending fluctuations over time after removing the school-averaged waveform, and the time-averaged separation between the center of mass of each swimmer.
The difference between the two network graphs in \Cref{fig:fish_swimming}e indicates that gait similarity exists among swimmers but is not intuitively related to spatial proximity, alluding to complex dynamical interactions that governs the emergent schooling behavior.

This study showcases that {\elastica}~can seamlessly integrate with external software, with little compromise on the model fidelity and computational efficiency of either software, laying groundwork for incorporating bio-locomotion and soft robotics in complex multi-physics environments.
Moreover, the large-scale FSI problem demonstrates how slender structures modeled in {\elastica}~can easily be distributed using message passing interface (MPI) across distributed memory when inter-rod coupling is minimal, which exploits an additional level of parallelism.

%% file: 03_discussion.tex
\section*{Discussion}
We have presented \elastica, a high-performance, open-source computational framework for simulating the dynamics of soft, slender structures based on the Cosserat rod theory. 
Designed with both modeling accuracy and computational efficiency in mind, \elastica~offers the versatility and performance for a wide range of applications without compromising the fidelity of the underlying continuum physics. 
Leveraging a series of HPC techniques, \elastica~exploits parallelism at the instruction, data, and thread levels to achieve significant performance gains over prior implementations, enabling large-scale simulations of heterogeneous filament assemblies. 
Moreover, we have demonstrated the software to be robust, flexible, and scalable across a variety of physical settings spanning mechanical systems, soft robotics, and emergent phenomena.
Indeed, \elastica~efficiently handles connections and frictional contacts between filaments and external boundaries, as well as the various modes of actuation including self-induced mechanical moments, gravitational and magnetic fields, and hydrodynamic interactions facilitated by third-party software. 
Therefore, we deliver the \elastica~framework as a powerful computational tool for modeling and simulating the dynamics of soft filaments in diverse application settings, with practical implications across multiple fields including biophysical research, robotic design, and biomedical engineering. 

%% file: 04_methods.tex
\section*{Methods}
\label{sec:methods}

\subsection*{Cosserat rod dynamics}

The Cosserat rod theory, a subclass of the Cosserat continuums, provides a one-dimensional geometrically exact, non-linear description of slender bodies undergoing large deformations in three-dimensional space~\cite{antman2005nonlinear,gazzola2018forward}.
A Cosserat rod deforming in 3D-space is described by a material curve $\mathcal{C}$ with a material curvilinear coordinate $\hat{s} \in [0, L_0]$, which is also the arc length parameter in the reference configuration. 
This curve defines the centerline of the rod, the position of which may be parameterized by $\bv{r}(\hat{s}, t) \in \mathbb{R}^3$ where $t \geq 0$ is the temporal variable. 
Upon each point on this curve is attached a triad of time-dependent orthonormal directors $\bv{d}_1(\hat{s}, t), \bv{d}_2(\hat{s}, t), \bv{d}_3(\hat{s}, t)$, which forms a proper orthogonal row matrix $\bt{Q}(\hat{s}, t) = {\{\bv{d}_1, \bv{d}_2, \bv{d}_3\}}^T$, which we call the director matrix. 
For any vector $\bv{v}$ defined in the Laboratory frame, its counterpart in the local frame is denoted $\bar{\bv{v}} = \bt{Q}\bv{v}$. 
Hence, the appropriate strain measures consist of $\bar{\bv{\sigma}} = \bt{Q}(\bv{r}' - \bv{d}_3)$, which encodes shearing and stretching, and $\bar{\bv{\kappa}} = \textup{ax}\left( \bt{Q} \partial_{\hat{s}} \bt{Q}^T \right)$ which is the generalized curvature vector measuring flexure and twisting. Here, the prime symbol $(\cdot)'$ denotes the partial derivative with respect to the material coordinate $\hat{s}$, and $\textup{ax}(\cdot)$ is the axis-vector operator defined as $\textup{ax}(\bt{U}) = \epsilon_{ijk} U_{kj} / 2$. 
Furthermore, we define the local frame angular velocity $\bar{\bv{\omega}} = \textup{ax}(\bt{Q} \partial_t\bt{Q}^T)$, material density per unit length $\rho$, cross-sectional area $A$, Young's ($E$) and shear ($G$) moduli, shear correction coefficient $\alpha_c$, axial dilatation $e = \|\bv{r}'\|$, elasticity tensors $\bt{B} = \textrm{diag}\{EI_1, EI_2, G(I_1+I_2)\}$ (bend-twist) and $\bt{S} = \textrm{diag}\{\alpha_c GA, \alpha_c GA, EA\}$ (shear-stretch), and inertia tensor $\bt{I} = \textrm{diag}\{I_1, I_2, I_1+I_2\}$. 
The governing equations (linear and angular momentum balance) for the rod dynamics then read
\begin{subequations}
\begin{equation}\label{eqn:elastica_lin_mom}
    \rho A \frac{\partial^2 \bv{r}}{\partial t^2} = \frac{\partial}{\partial \hat{s}}\left(\frac{\bt{Q}^T \bt{S}\bar{\bv{\sigma}}}{e}\right) + \bv{f},
\end{equation}
\begin{equation}\label{eqn:elastica_ang_mom}
\begin{aligned}
    \frac{\rho \bt{I}}{e} \frac{\partial \bar{\bv{\omega}}}{\partial t} &= \frac{\partial}{\partial \hat{s}}\left(\frac{\bt{B}\bar{\bv{\kappa}}}{e^3}\right) + \frac{\bar{\bv{\kappa}} \times \bt{B}\bar{\bv{\kappa}}}{e^3} + \bt{Q}\bv{t} \times \bt{S}\bar{\bv{\sigma}} \\
    &~~~+ \frac{\rho \bt{I}\bar{\bv{\omega}}}{e} \times \bar{\bv{\omega}} + \frac{\rho \bt{I}\bar{\bv{\omega}}}{e^2} \frac{\partial e}{\partial t} + \bar{\bv{c}},
\end{aligned}
\end{equation}
\end{subequations}
where $\bv{f}$ and $\bar{\bv{c}}$ represent the external distributed forces and couples. 
This representation entails a number of favorable features: 
\textbf{(1)} it captures 3D dynamic effects and all modes of deformation (bending, twist, shear, stretch); 
\textbf{(2)} distributed actuation and environmental coupling are directly attained via $\bv{f}$ and $\bar{\bv{c}}$, rendering the inclusion of (self-)contact, friction or electromagnetic effects straightforward; 
\textbf{(3)} complexity scales $\order{n}$ with axial resolution; and 
\textbf{(4)} kinematic/dynamic boundary conditions can be directly enforced, allowing to assemble multiple rods into complex layouts.
Additional details on the analytical and computational modeling of Cosserat rods, along with numerical solution algorithms, can be found in supplementary materials, Section 1. 

\subsection*{Magnetized filaments}
\label{subsec:magnetized-filaments}

Programmable magnetic cilia carpets~\cite{gu2020magnetic,dong2020bioinspired} are represented by extending Cosserat rod model: each filament carries a prescribed magnetization $\bar{\bv{m}}$ in the local director frame $\bt{Q}$, which couples to an imposed, time-varying laboratory magnetic field $\bv{b}(t)$.
For a spatially homogeneous $\bv{b}(t)$, the magnetic torque enters the angular momentum \Cref{eqn:elastica_ang_mom} through the external couple
\begin{equation}\label{eqn:magnetic_couple}
    \bar{\bv{c}}_{\mathrm{mag}} = \bar{\bv{m}} \times \left(\bt{Q}\bv{b}(t)\right),
\end{equation}
which is accumulated into $\bar{\bv{c}}$ together with any other distributed couples acting on the rod.

%% file: 05_acknowledgements.tex

\subsection*{Code availability}
The source code and all simulation scripts are publicly available at https://github.com/GazzolaLab/elasticapp. 

\subsection*{Acknowledgments}
This study was jointly funded by NSF EFRI C3 SoRo \#1830881, NSF CAREER \#1846752, NSF ELEMENTS \#2209322, ONR MURI \#N00014--19--1--2373, and the Strategic Research Initiatives (SRI) program of University of Illinois Urbana Champaign.
This work used Bridges-2 at Pittsburgh Supercomputing Center as well as Expanse at San Diego Supercomputing Center, through allocation \#MCB190004, \#BIO240096, and \#MCH250053 from ACCESS program.

\subsection*{Author contributions statement}
T.P. and S.H.K. designed, developed, and tested the initial software infrastructure. 
S.C. developed and tested the extended features related to third-party software integration. 
T.P., S.H.K., and S.C. implemented and performed the case studies and data analysis. 
M.G. conceptualized and supervised the project. 
T.P. provided the initial draft, and all authors contributed to the final manuscript. 


\subsection*{Materials \& Correspondence}
Correspondence should be addressed to M.G.